# Speaker verification using attentive multi-scale convolutional recurrent network

Yanxiong Li[*], Zhongjie Jiang, Wenchang Cao, Qisheng Huang

School of Electronic and Information Engineering, South China University of Technology, Guangzhou, China

**Abstract**

In this paper, we propose a speaker verification method by an Attentive Multi-scale Convolutional Recurrent Network (AMCRN). The proposed AMCRN can acquire both local spatial information and global sequential information from the input speech recordings. In the proposed method, logarithm Mel spectrum is extracted from each speech recording and then fed to the proposed AMCRN for learning speaker embedding. Afterwards, the learned speaker embedding is fed to the back-end classifier (such as cosine similarity metric) for scoring in the testing stage. The proposed method is compared with state-of-the-art methods for speaker verification. Experimental data are three public datasets that are selected from two large-scale speech corpora (VoxCeleb1 and VoxCeleb2). Experimental results show that our method exceeds baseline methods in terms of equal error rate and minimal detection cost function, and has advantages over most of baseline methods in terms of computational complexity and memory requirement. In addition, our method generalizes well across truncated speech segments with different durations, and the speaker embedding learned by the proposed AMCRN has stronger generalization ability across two back-end classifiers.

***Keywords:*** Speaker embedding, speaker verification, attentive mechanism, multi-scale convolutional recurrent network, dilated convolution

## 1. Introduction

Speaker recognition is a task that identifies a person based on his or her voices [1]. With the wide application of voice-enabled devices, especially smart phone, speaker recognition has become an indispensable component in many applications, such as criminal investigation [2], financial services [3]. For example, in the context of criminal investigation, law enforcement agencies often need to know whether the speech recordings are uttered by the claimed speaker or not. The law enforcement agencies want to use speaker recognition as a tool to investigate a suspect or to determine a judgment of guilt or innocence [2]. Therefore, speaker recognition is helpful for the law enforcement agencies to solve criminal cases [2]. In addition, speaker recognition is critical for successfully implementing other tasks related to multiple speakers, such as speaker diarization [4], speaker tracking [5], and multi-speaker speech recognition [6].

The task of speaker recognition can be divided into two categories: speaker identification and speaker verification [7]. Speaker identification is to decide which enrolled speaker utters a given voice from a set of known speakers [8]. Speaker verification is to reject or accept the identity claim of a speaker based on the speaker's utterance [9]. The work in this paper focuses on discussing speaker verification only.

*1.1. Related works*

Many works are done on speaker verification [10]-[25]. The efforts in these works mainly concentrate on two questions. The first question is how to learn (or extract) a discriminative front-end feature. The second question is how to build (or train) an effective back-end classifier.

---

[*] Corresponding author (Yanxiong Li, eeyxli@scut.edu.cn).

Many hand-crafted features are widely used to represent speaker's time-frequency properties, such as constant Q cepstral coefficients [26], Linear Prediction Coding Coefficients (LPCC) [1], Mel-Frequency Cepstral Coefficients (MFCC) [1], eigenvoice-motivated vectors [27], and I-vector [28]. Hand-crafted features are generally designed for specific situations. As a result, they lack generalization ability when they are adopted in other conditions. In addition, they are shallow-model based features. Therefore, they cannot deeply represent the differences of time-frequency properties among different speakers. To overcome the deficiencies of these hand-crafted features, some deep-model based features are learned by deep neural networks. For example, the X-vector is learned by a Time-Delay Neural Network (TDNN) [29]-[32], and the speaker embeddings learned by: a Convolutional Neural Network (CNN) [4], [33], [34], a Long Short-Term Memory Network (LSTMN) [35], a Siamese Neural Network (SNN) [36], or a Res2Net [37]. On the other hand, back-end classifiers that are commonly used in previous works, are Probabilistic Linear Discriminant Analysis (PLDA) [38] and Cosine Similarity Metric (CSM) [1].

For example, Chowdhury et al [1] approach the problem of speaker verification from severely degraded audio data by judiciously combining two commonly hand-crafted features: MFCC and LPCC. They first extract the features of both MFCC and LPCC from each speech recording. Afterwards, they use a 1D-Triplet-CNN to combine these two features for representing the characteristic differences among different speakers. Finally, a back-end classifier of CSM is adopted to score on the testing speech recordings. Experimental results reveal that their method is robust to a wide range of audio degradations. Another representative work for speaker verification is done by Snyder et al [30], in which they propose a method for learning speaker embedding using the TDNN. In their system, the front-end feature is the X-vector and the back-end classifier is the PLDA. Their experimental results demonstrate that their proposed system significantly outperforms two baseline systems based on the I-vector.

*1.2. Our contributions*

In the speaker verification systems based on deep neural network, some deep models are used for speaker embedding learning, such as TDNN, CNN, LSTMN, SNN, and Res2Net. They obtain state-of-the-art results and exceed shallow-model based methods, such as the I-vector based method [39]. However, there is still room for improvement in these methods. First, the deep models adopted in previous works are either single-scale convolutional-block based network (such as CNN, Res2Net) or sequential-block based network (such as TDNN, LSTMN). The single-scale convolutional-block based networks are skilled in capturing local spatial information, while sequential-block based networks are good at acquiring global sequential information [40]. Therefore, it is quite difficult for the speaker embedding learned by previous deep models to represent both local spatial information and global sequential information contained in the speech samples. Second, multi-scale convolutional block with temporal attention is not explicitly considered in previous networks, which is able to capture local spatial information with multiple resolutions. In addition, both local spatial information and global sequential information are important for learning a speaker embedding with strong ability to represent the characteristic differences among different speakers. As a result, the ability of previous deep models for speaker embedding learning is limited, and the discriminative information among different speakers is insufficiently learned.

We design a deep model of AMCRN to learn the speaker embedding which can effectively represent both local spatial information and global sequential information. We propose a method of speaker verification using the AMCRN. In our method, audio feature is first extracted from each speech sample and fed to the AMCRN for learning speaker embedding. Afterwards, the learned speaker embedding is fed to the classifier of CSM or PLDA for scoring in the testing stage. Different methods are assessed on three public datasets which are selected from two large-scale speech corpora: VoxCeleb1 and VoxCeleb2. In addition, we discuss the impacts of the AMCRN's structures on the performance of the proposed method, and the generalization ability of our method across truncated speech segments. Experimental results show that the proposed method achieves satisfactory results.

In summary, main contributions of this study are as follows:

First, we propose a deep model of AMCRN for learning a speaker embedding with strong ability to represent the characteristic differences among different speakers. Main components of the AMCRN consist of two types of blocks. The first one is the multi-scale convolutional block with temporal attention, which is a unique part of the AMCRN. The second one is the sequential (recurrent) block. As a result, the proposed AMCRN can acquire both local spatial information with multiple resolutions and global sequential information from each input speech sample.

Second, we propose a speaker verification method based on the AMCRN. The effectiveness of the proposed method is evaluated from multiple aspects on three public datasets. Experimental results show that our method outperforms state-of-the-art methods in terms of both Equal Error Rate (EER) and Minimal Detection Cost Function (minDCF). In addition, our method has advantages over previous methods in terms of computational complexity and memory requirement.

The rest of the paper is structured as follows. Section 2 describes the proposed method in detail. Section 3 presents the experiments and discussions. Finally, conclusions and future work are given in Section 4.

## 2. Method

Fig. 1 illustrates the framework of the proposed speaker verification method using the AMCRN. The AMCRN is composed of one Speaker Embedding Module (SEM), and one Output Module (OM). The SEM consists of one initial convolutional layer, some multi-scale convolutional blocks with temporal attention, one residual Bidirectional Long Short-Term Memory (BLSTM) block, one channel attentive statistic pooling layer, and one fully connected layer with batch normalization. The OM is one classifier of Additive Angular Margin (AAM) Softmax [41].

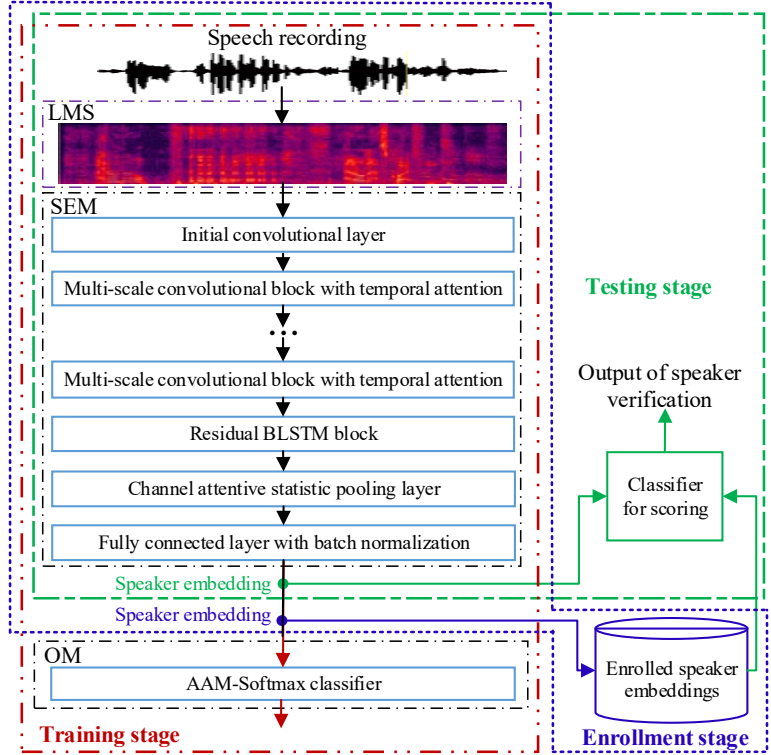

**Fig. 1.** The framework of the proposed speaker verification method using the AMCRN. The contents in red dot-dashed, blue dotted and green dashed boxes are conducted in the training, enrollment and testing stages, respectively.

The SEM is proposed to learn speaker embedding with strong ability for representing characteristic differences of various speakers. The motivation for designing the SEM is based on the following three considerations.

First, we propose a multi-scale convolutional block with temporal attention for acquiring local spatial information with several resolutions from speech recordings. The multi-scale convolutional block has the ability to acquire local spatial information with multiple resolutions contained in each feature map of speech recordings. The local spatial information can represent the detailed differences of time-frequency properties among various speakers. In addition, the multi-scale convolutional block is a residual structure which can ease the training of the AMCRN and exceeds single-scale convolutional structure for the tasks of recognition [42], [43]. The temporal attention can make the AMCRN concentrate on critical speech frames of each speech recording in acquiring local spatial information. In a speech recording, the critical speech frames have greater contributions for acquiring discriminative information of speakers than other speech frames.

Second, we design a global sequential block (i.e., residual BLSTM block) for acquiring global sequential information of a speech recording. That is, the SEM is a multi-scale Convolutional Recurrent Neural Network (CRNN) which consists of some multi-scale convolutional blocks (namely CNN) and one residual BLSTM block (namely one Recurrent Neural Network (RNN)). What's more, CRNN outperforms both CNN and RNN for other audio processing tasks [44], [45].

Third, the channel attentive statistic pooling layer and the fully connected layer with batch normalization are adopted to project the output of the residual BLSTM block into a fixed-length supervector (i.e., speaker embedding). Hence, the learned speaker embedding is expected to be an effective and compact representation for each input speech recording.

As depicted in Fig. 1, the work procedures of the proposed method at different stages are as follows. In the training stage (the contents in red dot-dashed box), the AMCRN is trained to learn speaker embeddings from the Logarithm Mel Spectra (LMS) of the training speech recordings. In the enrollment stage (the contents in blue dotted box), speaker embeddings are generated by the SEM of the AMCRN from the LMS of the enrollment speech recordings. In the testing stage (the contents in green dashed box), the speaker embedding of each testing speech recording is outputted from the SEM of the AMCRN. Afterwards, the similarity between each testing speaker embedding and the enrolled speaker embedding corresponding to the claimed speaker is calculated by a classifier of CSM or PLDA. If the similarity is lower than a threshold (e.g., 0.5), then the claimed speaker identity of the testing speech recording is rejected; otherwise, it is accepted.

*2.1. LMS extraction*

LMS has been successfully used as input audio feature of deep neural network to learn deep embedding [44], [46]. The block diagram of the LMS extraction is illustrated in Fig. 2. The LMS extraction includes four steps: Hamming windowing, square of absolute value of Fast Fourier Transformation (FFT), Mel-scale filtering and logarithm operation. The procedure for extracting the LMS is described in detail as follows. First, each speech recording is split into overlapping frames and windowed by a Hamming window. Afterwards, the FFT is performed on the windowed frames to produce power spectrum which is then smoothed by a bank of Mel-scale filters. Finally, logarithm operation is implemented on the output of Mel-scale filters to obtain the LMS feature.

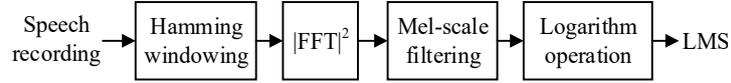

**Fig. 2.** The extraction procedure of LMS, which include steps of Hamming windowing, square of absolute value of FFT, Mel-scale filtering and logarithm operation.

*2.2. Speaker embedding module*

A SEM is proposed to learn speaker embedding from the LMS of each speech recording. The structure of the SEM is depicted in Fig. 1. Each component of the proposed SEM is introduced as follows.

*2.2.1. Multi-scale convolutional block with temporal attention*

Fig. 3 illustrates the proposed multi-scale convolutional block with temporal attention. It mainly includes the following operations: dilated convolution with batch normalization and Rectified Linear Unit (ReLU), dilated Conv 3, dilated convolution with batch normalization, temporal attention, element-wise multiplication, element-wise summation, and activation function of ReLU. Here, dilated Conv 3 represents one-dimensional dilated convolution with convolutional kernel size of 3. The motivation for designing the multi-scale convolutional block is that the multi-scale convolutional block can learn more detailed local spatial information compared to the single-scale convolutional block. In addition, dilated convolution instead of standard convolution is adopted in the proposed multi-scale convolutional block. The consideration is that dilated convolution has larger receptive field to acquire longer contextual information without increasing the amount of network's parameters. What's more, it has obtained more satisfactory results than standard convolution in other audio processing tasks [44].

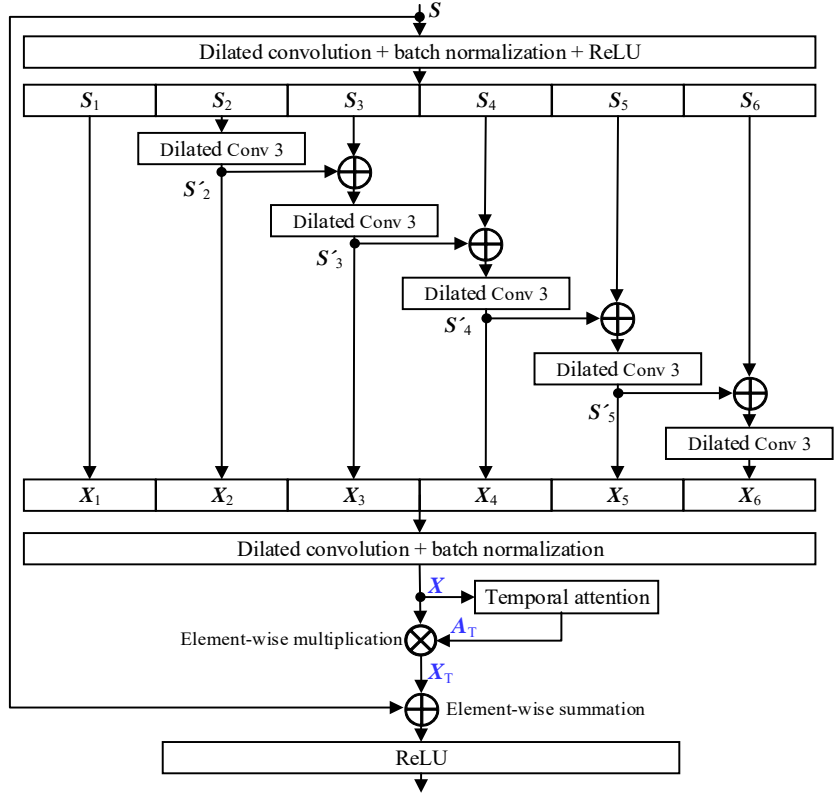

**Fig. 3.** The proposed multi-scale convolutional block with temporal attention, where six-scale is taken as an example for the convenience of description.

For the convenience of description, we take six-scale (six feature subsets) as an example as shown in Fig. 3. The optimal setting of the scale number will be discussed in sub-section 3.3.1. The work procedure of multi-scale convolutional block is as follows. First, the input feature map $S$ is fed to a layer of dilated convolution followed by the operations of batch normalization and ReLU. Second, the feature map $S$ is split into six feature subsets, $S_1$ to $S_6$. That is, feature map $S$ is processed in a manner of multi-scale (e.g., six-scale in Fig. 3), which is of benefit for learning local spatial information with multiple resolutions. Each feature subset $S_i$ ($1 \leq i \leq 6$) has the same spatial size with 1/6 number of channels compared to the feature map $S$. The operation of dilated Conv 3 is implemented on the feature subsets $S_2$ to $S_6$ for further transformation. To reduce parameter size of the multi-scale convolutional block and increase the diversities of feature subsets, dilated Conv 3 is not conducted on the feature subset $S_1$. In addition, the element-wise summations of feature subsets: $S_3$ and $S'_2$, $S_4$ and $S'_3$, $S_5$ and $S'_4$, $S_6$ and $S'_5$, are performed before the operation of dilated Conv 3. The consideration for implementing the element-wise summation is that each dilated Conv 3 for $S_i$ ($3 \leq i \leq 6$) receives discriminative information from feature subsets $S_j$ ($j<i$) and there are information interactions between adjacent feature subsets. These information interactions are beneficial for improving the representation ability of the learned speaker embedding. Third, to fuse discriminative information from different scales, all feature subsets are first concatenated, and their concatenation is fed to a layer of dilated convolution followed by the operation of batch normalization. Fourth, inspired by the success of attention mechanism to the computer vision [47], temporal attention is integrated into the multi-scale convolutional block. As a result, the proposed SEM can concentrate on important frames of each speech recording in capturing local spatial information. Fifth, the feature map $S$ is element-wisely summed with the feature map $X_T$. Finally, the result of element-wise summation is fed to a layer of ReLU for outputting the final feature map.

To acquire longer time-frequency contextual information with larger receptive field, dilated convolutions instead of standard convolutions are exploited in the proposed SEM. A two-dimensional convolution $*$ [44], is defined by

$$(F * K)(q) = \sum_{s+t=q} F(s)K(t), \qquad (1)$$

where $F$ stands for the input feature; $K$ denotes convolutional kernel whose size is $(2n+1)\times(2n+1)$; $q$, $s \in \mathbb{Z}^2$, $t \in [-n, n]^2 \cap \mathbb{Z}^2$, and $\mathbb{Z}$ is the set of integers. The dilated convolutional $*_r$, is defined by

$$(F *_r K)(q) = \sum_{s+rt=q} F(s)K(t), \qquad (2)$$

where $q$, $s \in \mathbb{Z}^2$, $t \in [-n, n]^2 \cap \mathbb{Z}^2$, $r$ is a dilation rate, and $*_r$ is called a dilated convolution with dilation rate $r$. In this sense, a standard convolution can be regarded as a special dilated convolution with dilation rate 1. A one-dimensional dilated convolution with dilation rate $r$ can be also defined by Eq. (2) after changing the definition domain of the variables. That is, in the one-dimensional dilated convolution, the variables meet the following requirements: $q$, $s \in \mathbb{Z}$, $t \in [-n, n] \cap \mathbb{Z}$. In addition, the stride of convolutions is set to 1 for keeping the time resolution the same as in the input.

The temporal attention is realized by exploiting the inter-time relationship of input feature map at the frame level. The temporal attention can make the SEM focus on critical speech frames of the input feature map. The detailed procedure for temporal attention operation is illustrated in Fig. 4.

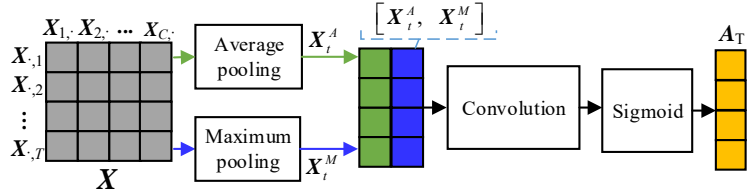

**Fig. 4.** The illustration for temporal attention operation, which consists of the blocks of average pooling, maximum pooling, concatenation, convolution and sigmoid.

Let the input feature map be $\mathbf{X} \in \mathbf{R}^{T \times C}$, where $T$ and $C$ represent total number of frames and channels, respectively. Average pooling operation and maximum pooling operation are performed on the channel dimension of feature map. Let the outputs of average pooling operation and maximum pooling operation be $\mathbf{X}_t^A = \{X_t^A\} \in \mathbf{R}^{T \times 1}$ and $\mathbf{X}_t^M = \{X_t^M\} \in \mathbf{R}^{T \times 1}$, respectively, where $t$ is frame index and $1 \leq t \leq T$. The definitions of $X_t^A$ and $X_t^M$ are defined by Eq. (3) and Eq. (4), respectively.

$$X_t^A = \frac{1}{C}\sum_{c=1}^{C} X_t(c), \qquad (3)$$

$$X_t^M = \max\left(X_t(1),\ X_t(2),\ ...,\ X_t(C)\right). \qquad (4)$$

Next, $X_t^A$ and $X_t^M$ are concatenated and then an operation of convolution is carried out on their concatenation. Afterwards, the output of the operation of convolution is used as the input of a Sigmoid function for producing the coefficient vector $\mathbf{A}_T$. That is, $\mathbf{A}_T$ is defined by

$$\mathbf{A}_T = \varphi\left(Conv\left(\left[\mathbf{X}_t^A,\ \mathbf{X}_t^M\right]\right)\right), \qquad (5)$$

where $\varphi(\cdot)$ and $Conv(\cdot)$ denote the Sigmoid function and the convolutional operation, respectively; and $[\mathbf{X}_t^A, \mathbf{X}_t^M] \in \mathbf{R}^{T \times 2}$ stands for the concatenation of $\mathbf{X}_t^A$ and $\mathbf{X}_t^M$. Finally, the output of temporal attention, $X_T$, is generated by conducting element-wise multiplication of both the input feature map $X$ and the coefficient vector $A_T$:

$$X_T = A_T \otimes X . \tag{6}$$

*2.2.2. Residual BLSTM block*

As illustrated in Fig. 1, the input feature map of residual BLSTM block is the output feature map of the last multi-scale convolutional block. The residual BLSTM block consists of two BLSTM blocks and one fully connected layer, which is depicted in Fig. 5.

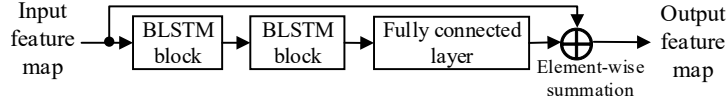

**Fig. 5.** The framework of the residual BLSTM block, which consists of two BLSTM blocks and one fully connected layer.

The BLSTM block which is a combination of a Long Short-Term Memory Module (LSTMM) and a Bidirectional Recurrent Module (BRM) [48]. The LSTMM can effectively utilize the contextual information for long periods of time. However, it gains the contextual information from forward direction only and cannot have access to the contextual information from backward direction. As for the task of speaker verification, we need to adopt the contextual information from two directions. The BRM can implement this target by processing the speech recording with two separate hidden layers in forward direction and backward direction [49]. As a result, the BLSTM block can utilize the contextual information for long periods of time. The BLSTM has access to the contextual information in two directions with two separate hidden layers which are linked to the same output layer [50], as illustrated in Fig. 6.

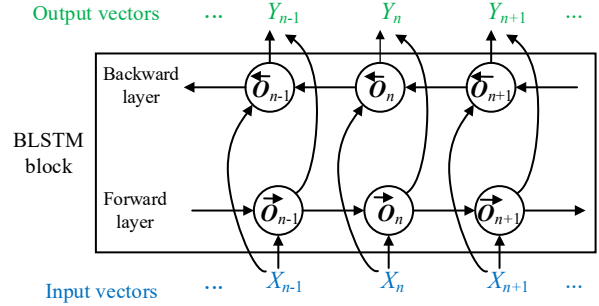

**Fig. 6.** The illustration for one BLSTM block.

The BLSTM block is expressed by

$$\begin{cases} \vec{O}_n = \Gamma\left(W_{\vec{I}} X_n + W_{\vec{O}} O_{n-1} + b_{\vec{O}}\right) \\ \overleftarrow{O}_n = \Gamma\left(W_{\overleftarrow{I}} X_n + W_{\overleftarrow{O}} O_{n+1} + b_{\overleftarrow{O}}\right), \\ Y_n = W_{\vec{Y}} \vec{O}_n + W_{\overleftarrow{Y}} \overleftarrow{O}_n + b_Y \end{cases} \tag{7}$$

where $X_n$ and $Y_n$ denote the inputs and outputs of the BLSTM block, respectively. $W$ and $b$ denote the weight matrix and bias vector, respectively. $\vec{O}_n$ and $\overleftarrow{O}_n$ are the forward and backward hidden sequence, respectively. $\Gamma$ is an activation function.

*2.2.3. Channel attentive statistic pooling and fully connected layers*

Inspired by the success of channel attention in the TDNN-based deep models for speaker verification [29], [51], we integrate a channel attentive statistic pooling layer into the proposed SEM. As illustrated in Fig. 1, the channel attentive statistic pooling layer is adopted to project the output of the residual BLSTM block into a compressed supervector. Afterwards, the supervector is fed to the fully connected layer with batch normalization for producing the speaker embedding. The detailed introduction to the channel attentive statistic pooling is presented in [29].

*2.3. Output module*

As shown in Fig. 1, the output module consists of one layer of AAM-Softmax classifier. In the training stage, speaker embedding learned by the SEM is fed to the output module for implementing the training of the AMCRN. However, the output module is not utilized in both the enrollment stage and the testing stage.

## 3. Experiments

In this section, we first describe experimental datasets and setups. Second, we discuss the ablation experiments on the settings of the SEM's structures. Third, we compare the proposed method with state-of-the-art methods on the experimental datasets. Fourth, we evaluate the robustness of different methods across truncated speech segments with various durations. Finally, we discuss the computational complexities and memory requirements of different methods.

*3.1. Experimental datasets*

Experimental datasets are chosen from two large-scale speech corpora: the VoxCeleb1 [52] and the VoxCeleb2 [53]. They are two of the most popular speech corpora that are adopted for speaker recognition in previous works. They are publicly available for research purposes.

The VoxCeleb1 has 153516 speech recordings uttered by 1251 speakers. Total duration of the VoxCeleb1 is 352 hours. All speech recordings of the VoxCeleb1 are from the sound tracks of YouTube videos. The dataset is basically gender balanced, with 55% of male speakers. Speakers in the speech recordings span a great range of various ethnicities, professions, ages, and accents. The contents of the speech recordings mainly consist of interviews, speeches given to large audiences, outdoor stadiums and indoor studios, excerpts from professionally shot multimedia. There are some types of real-world noises in the speech recordings, including background utterances (chatter, laughter), noises of recording equipment, room reverbs.

The VoxCeleb2 is an extension of the VoxCeleb1, which has 1128246 speech recordings uttered by 6112 speakers. Total duration of the VoxCeleb2 is 2442 hours. The VoxCeleb2 dataset is roughly gender balanced, with 39% of female speakers. The speakers span a wide range of different ethnicities, accents, professions, and ages. Speech recordings of the dataset are degraded with background laughter, chatter, overlapping speech, and varying room acoustics.

Experimental data are selected from both VoxCeleb1 and VoxCeleb2, and include three public datasets: VoxCeleb2-dev, VoxCeleb1-dev, and VoxCeleb1-test. These three datasets are divided by the data organizer of VoxCeleb [54] and popularly used in previous works [52], [53]. Training data are selected from VoxCeleb2-dev, and have 5994 speakers with 1092009 speech recordings in total. Testing data are chosen from both VoxCeleb1-dev and VoxCeleb1-test, and consist of three data subsets: Vox1-O, Vox1-E and Vox1-H. These data subsets are divided by the data organizer of VoxCeleb [54]. O, E, and H in the names of these three data subsets represent original dataset, extended dataset, and hard dataset of VoxCeleb1, respectively. The proportions of training subset and validation subset in the training data are 95% and 5%, respectively. The average length of speech recordings is approximately 8 s. The detailed information of experimental data is listed in Table 1.

**Table 1**

The detailed information of experimental data. The training data are the development dataset of VoxCeleb2 that consists of 5994 speakers with 1092009 recordings. The testing data include three data subsets: Vox1-O (37611 trails from 40 speakers), Vox1-E (579818 trails from 1211 speakers) and Vox1-H (550894 trails from 1211 speakers).

|  | Names | Dataset sources | #Speakers | Data amount |
|---|---|---|---|---|
| Training | / | VoxCeleb2-dev | 5994 | 1092009 recordings |
| Testing | Vox1-O | VoxCeleb1-test | 40 | 37611 trails |
|  | Vox1-E | VoxCeleb1-dev | 1211 | 579818 trails |
|  | Vox1-H | VoxCeleb1-dev | 1211 | 550894 trails |

*3.2. Experimental setup*

All experiments are conducted on a server computer whose main configurations are as follows: an Intel CPU i7-6700 with 3.10 GHz, a RAM of 64 GB, and a NVIDIA 1080TI GPU. The baseline and the proposed methods are implemented on the PyTorch toolkit. Two most popular metrics: EER and minDCF (P-target = 0.01), are adopted to measure the performance of all methods. The detailed definitions of both EER and minDCF are referred to [55]. The lower the scores of the EER and the minDCF are, the better the performance of the speaker verification method is.

The input audio feature of the AMCRN is the 80-dimensional LMS which is extracted from each speech recording. The LMS is subjected to short time cepstral mean variance normalization with a sliding window of 3 s. To increase the diversity of training data, we conduct the operation of data augmentation on the training data in all experiments. Each training (i.e., original) speech recording is used to generate two copies of augmented speech recordings by randomly adding background noise, music, babble, and reverb on the original speech recording. Both original speech recordings and augmented speech recordings are used in the experiment with data augmentation. The configurations of main parameters of the proposed method are presented in Table 2.

**Table 2**

Configurations of main parameters of the proposed method. The setting of the values of main parameters of our method is a conventional practice or determined based on the training data.

| Type | Parameters' settings |
|---|---|
| Preprocessing | Frame length: 25 ms |
|  | Frame shift: 10 ms |
|  | Dimension of LMS: 80 |
| SEM | Batch size: 256 |
|  | Learning rate: 0.005 to 0.000001 |
|  | Dropout ratio of BLSTM: 0.2 |
|  | Kernel size of initial convolutional layer: 5 |
|  | Number of Multi-scale Convolutional Block (MCB): 3 |
|  | Number of channels in each MCB: [512, 512, 512] |
|  | Kernel size of each MCB: [3, 3, 3] |
|  | Dilation rates in each MCB: [2, 3, 4] |

| | Output channel number per MCB: [512, 512, 512] |
| | Number of BLSTM blocks in the residual BLSTM block: 2 |
| | Number of shared neurons in a BLSTM cell: 450 |
| | Number of channels in pooling layer: 128 |
| | Dimension of speaker embedding: 256 |
| | Neurons of the fully connected layer: 256 |
| OM | Margin of AAM-Softmax: 0.2 |
| | Pre-scaling of AAM-Softmax: 30 |

The AMCRN's parameters are optimized on the training data via the Adam optimizer with binary cross-entropy loss [56]. After each epoch on the training subset, the trained AMCRN is evaluated on the validation subset. The AMCRN with the lowest validation loss is adopted as the final network. After finishing the network training, enrollment data and testing data are fed to the SEM for learning speaker embeddings. Finally, similarity between the testing speaker embedding and the corresponding enrolled speaker embedding is measured by the back-end classifiers of CSM or PLDA for scoring.

*3.3. Ablation experiments on the SEM*

In this sub-section, we first discuss the configurations of scale's number of multi-scale convolutional block, and then discuss the optimal settings of the SEM's structures. In the experiments of this sub-section, the back-end classifier of CSM is adopted for scoring.

*3.3.1. Settings of scale's number of multi-scale convolutional block*

We compare the performance of the AMCRN with different scale's numbers of the multi-scale convolutional block as shown in Table 3. The number of scales is variable, whereas other parts of the SEM is set to: dilated convolutional block with temporal attention and residual BLSTM block. The scores of both EER (in %) and minDCF obtained by the AMCRN with different scale's numbers on testing data are listed in Table 3. Our method obtains the lowest scores of both EER and minDCF, when the number of scales is set to 8. When the number of scales of multi-scale convolutional block deviates from 8, the scores of both EER and minDCF steadily increase. Hence, the number of scales is set to 8 in the experiments of the following sections.

**Table 3**

Impacts of the scale's numbers on the performance of the AMCRN for speaker verification on experimental data. The performance metrics are EER (in %) and minDCF.

| No. of scales | Vox1-O (EER/minDCF) | Vox1-E (EER/minDCF) | Vox1-H (EER/minDCF) |
|---|---|---|---|
| 2 | 1.527/0.188 | 1.702/0.205 | 2.846/0.312 |
| 4 | 1.469/0.180 | 1.560/0.182 | 2.679/0.293 |
| 8 | **1.464/0.177** | **1.551/0.182** | **2.640/0.284** |
| 16 | 1.571/0.193 | 1.7858/0.214 | 2.934/0.330 |
| 32 | 1.656/0.192 | 2.018/0.237 | 3.160/0.356 |

*3.3.2. Settings of the SEM's structures*

We compare the performance of the AMCRN with different structures of the SEM as shown in Table 4. For example, the structure of "Ms-Dil-Conv with TA + RBLSTM" denotes a structure of the SEM which consists of multi-scale dilated convolutional blocks with temporal attention and residual

BLSTM block. The scores of both EER (in %) and minDCF obtained by the AMCRN with different structures of the SEM on testing data are listed in Table 4.

**Table 4**

Impacts of the SEM's structures on the performance of the AMCRN for speaker verification on experimental data. The performance metrics are EER (in %) and minDCF.

| No. | Structures of the SEM | Vox1-O (EER/minDCF) | Vox1-E (EER/minDCF) | Vox1-H (EER/minDCF) |
|---|---|---|---|---|
| ① | Ms-Dil-Conv with TA + RBLSTM | **1.464/0.177** | **1.551/0.182** | **2.640/0.284** |
| ② | Ss-Dil-Conv with TA + RBLSTM | 1.711/0.198 | 1.783/0.210 | 2.933/0.318 |
| ③ | Ms-Sta-Conv with TA + RBLSTM | 1.488/0.179 | 1.598/0.183 | 2.736/0.289 |
| ④ | Ms-Dil-Conv + RBLSTM | 1.502/0.182 | 1.619/0.183 | 2.743/0.301 |
| ⑤ | Ms-Dil-Conv with TA | 1.657/0.190 | 1.742/0.203 | 2.913/0.312 |

Ms-Dil-Conv: multi-scale dilated convolution; RBLSTM: residual BLSTM; TA: temporal attention; Ms-Sta-Conv: multi-scale standard convolution; Ss-Dil-Conv: single-scale dilated convolution; EER is in %.

When the SEM's structure is set to "Ms-Dil-Conv with TA + RBLSTM", our method reaches the lowest EER scores of 1.464% on Vox1-O, 1.551% on Vox1-E, and 2.640% on Vox1-H. Similarly, our method obtains the lowest minDCF scores of 0.177 on Vox1-O, 0.182 on Vox1-E, and 0.284 on Vox1-H. Higher scores of both EER and minDCF are obtained, when the SEM is set to other structures as listed in Table 4. Hence, the speaker embedding learned by the SEM with "Ms-Dil-Conv with TA + RBLSTM" outperforms the deep embeddings that are learned by the SEM with other structures. The proposed SEM's structure mainly consists of the following parts: dilated convolution, multi-scale convolutional block, temporal attention, and residual BLSTM block. Experimental results in Table 4 show that all components of the proposed SEM's structure contribute to the performance improvement of the proposed method. For example, when evaluated on the Vox1-H, 0.293%, 0.096%, 0.103%, and 0.273% of EER scores are decreased by introducing the multi-scale convolutional block (① versus ②), dilated convolution (① versus ③), temporal attention (① versus ④), and residual BLSTM block (① versus ⑤), respectively. Similarly, the MinDCF scores are also decreased by introducing these blocks above. In the experiments of the following sections, the structure of the proposed SEM is set to 8-scale dilated convolutional block with temporal attention and residual BLSTM block.

*3.4. Comparison of various methods*

In this sub-section, we compare the proposed method with the state-of-the-art methods for speaker verification. The first one is the TDNN based method, where speaker embedding (namely X-vector) is learned by the TDNN [30]. This method is proposed in 2018 and is one of the most representative deep-model-based methods for speaker verification. The second and third methods are the ResNet18 based and ResNet34 based methods, respectively. These two methods are proposed in 2019, where speaker embeddings are learned by the deep residual convolutional neural networks of ResNet18 and ResNet34, respectively [57]. The fourth one is the Res2Net based method proposed in 2021, in which speaker embedding is learned by the deep model of Res2Net [58]. The fifth one is the LSTMN based method proposed in 2018, in which speaker embedding is learned by the deep model of LSTMN [35]. The last one is the GAFAM (Graph Attentive Feature Aggregation Module) based method which is proposed in 2022 [59]. The objective of this method is to combine multiple frame-level features into a single utterance-level representation. A ResNet is adopted as the backbone network for learning speaker embedding in the experiments. Our method is the AMCRN based method. The back-end classifiers are CSM or PLDA for all methods.

Main parameters of the baseline methods are set based on the advisements in corresponding references and then tuned on the training data. Afterwards, our method and the baseline methods are evaluated on the testing data. The scores are not normalized when the CSM is used for scoring. We implement the TDNN based method by the codes in the Kaldi toolkit, whereas we re-implement other baseline methods based on the descriptions in corresponding references.

Under the same conditions, the scores of EER and minDCF that are obtained by different methods using the back-end classifier of CSM are presented in Table 5. Our method obtains the lowest EER scores of 1.464% on Vox1-O, 1.551% on Vox1-E, and 2.640% on Vox1-H. Similarly, our method achieves the lowest minDCF scores of 0.177 on Vox1-O, 0.182 on Vox1-E, and 0.284 on Vox1-H.

**Table 5**

Scores of EER (in %) and minDCF obtained by different methods using the back-end classifier of CSM.

| Methods | Vox1-O (EER/minDCF) | Vox1-E (EER/minDCF) | Vox1-H (EER/minDCF) |
|---|---|---|---|
| TDNN based | 6.327/0.679 | 6.449/0.681 | 9.103/0.995 |
| ResNet18 based | 1.936/0.225 | 2.120/0.247 | 3.794/0.413 |
| ResNet34 based | 1.674/0.193 | 1.752/0.195 | 3.440/0.389 |
| Res2Net based | 1.632/0.187 | 1.714/0.194 | 2.894/0.301 |
| LSTMN based | 2.245/0.267 | 2.288/0.275 | 3.966/0.441 |
| GAFAM based | 1.751/0.203 | 1.829/0.209 | 3.581/0.402 |
| AMCRN based | **1.464/0.177** | **1.551/0.182** | **2.640/0.284** |

Under the same experimental conditions, the scores of EER and minDCF that are obtained by different methods using the back-end classifier of PLDA are presented in Table 6. Our method produces the lowest EER scores of 1.739% on Vox1-O, 1.967% on Vox1-E, and 3.819% on Vox1-H. Similarly, our method obtains the lowest minDCF scores of 0.208 on Vox1-O, 0.237 on Vox1-E, and 0.401 on Vox1-H.

**Table 6**

Scores of EER (in %) and minDCF obtained by different methods using the back-end classifier of PLDA.

| Methods | Vox1-O (EER/minDCF) | Vox1-E (EER/minDCF) | Vox1-H (EER/minDCF) |
|---|---|---|---|
| TDNN based | 2.883/0.312 | 3.032/0.326 | 4.891/0.528 |
| ResNet18 based | 2.758/0.294 | 2.719/0.305 | 4.238/0.453 |
| ResNet34 based | 2.451/0.274 | 2.398/0.272 | 3.963/0.442 |
| Res2Net based | 2.411/0.263 | 2.354/0.273 | 3.920/0.421 |
| LSTMN based | 2.987/0.346 | 2.921/0.353 | 4.472/0.493 |
| GAFAM based | 2.538/0.284 | 2.475/0.286 | 4.104/0.449 |
| AMCRN based | **1.739/0.208** | **1.967/0.237** | **3.819/0.401** |

Based on the results in Tables 5 and 6, we can conclude that our method outperforms all baseline methods under the same conditions (scored by CSM or PLDA) in terms of EER and minDCF. The possible reason is that some measures are explicitly taken to learn speaker embedding in our method, such as dilated convolution, multi-scale convolutional block, temporal attention, and residual BLSTM block. These measures possess their own merits and work together to learn a speaker embedding with strong ability to represent the characteristic differences among different speakers. For instance, the multi-scale convolutional block captures local spatial information with several resolutions and the residual BLSTM block acquires global sequential information. Both local spatial information and global sequential information have contributions to the performance improvement of the speaker verification method. In addition, the following conclusions can be drawn.

First, when the back-end classifiers of CSM and PLDA are adopted for scoring, the scores of both EER and minDCF obtained by various methods are different. All methods, except the TDNN based method, achieve lower scores of EER and minDCF, when the CSM is used for scoring. On the contrary, the TDNN based method obtains lower scores of EER and minDCF, when the PLDA is adopted for scoring. Hence, the CSM is a more effective classifier in all methods, except the TDNN based method. The combination of front-end feature of X-vector and back-end classifier of PLDA is a better choice for the TDNN based method. This result of the TDNN based method is consistent with the previous practices in [30] and [31].

Second, the generalization capacity of different front-end features of speaker embeddings across various back-end classifiers are of difference. The absolute margins of both EER and minDCF scores achieved by the TDNN based method are the largest among all methods, when the classifiers of CSM and PLDA are adopted for scoring. The absolute margins of both EER and minDCF scores obtained by our method are the smallest among all methods, when the CSM and PLDA are used for scoring on the Vox1-O and Vox1-E. In addition, the absolute margins of both EER and minDCF scores obtained by the ResNet18 based method are the smallest, when the CSM and PLDA are adopted for scoring on the Vox1-H. The maximum absolute margins of EER and minDCF scores are 4.212% (9.103% - 4.891%) and 0.467 (0.995 - 0.528), respectively. These two maximum margins are obtained by the TDNN based method. Therefore, the generalization capacity of the X-vector is the worst among all front-end features of speaker embeddings, when different back-end classifiers are used to score. In contrast, the speaker embedding learned by the proposed AMCRN can generalize well across different back-end classifiers rather than overfitting one single classifier.

Third, all methods achieve the lowest scores of EER and minDCF on the Vox1-O, and produce the highest scores of both EER and minDCF on the Vox1-H. The possible reason is that the data size and complexity are of difference among these three testing data subsets. Specifically, the Vox1-O (Original dataset) is a relatively small-scale dataset with 37611 trails uttered by 40 speakers only. Therefore, the performance of all methods can be better when the Vox1-O is used as the testing data subset. Compared to the Vox1-O, both the Vox1-E (Extended dataset with 579818 trails) and the Vox1-H (Hard dataset with 550894 trails) are two large-scale datasets uttered by 1211 speakers. What's more, The Vox1-H is with the highest data complexity among these three testing data subsets.

*3.5. Performance comparison on truncated speech segments*

In this sub-section, we discuss the performance of different methods on the truncated testing speech segments with different durations. Based on the results in sub-section 3.4, it is known that the PLDA is a better choice for the X-vector, whereas the CSM is more suitable for other speaker embeddings. In this experiment, our method is denoted as: AMCRN based (CSM). The six baseline methods are denoted as follows: TNDD based (PLDA), ResNet18 based (CSM), ResNet34 based (CSM), Res2Net (CSM), LSTMN based (CSM), and GAFAM based (CSM). In the testing data, the average length of each testing speech recording (utterance) is approximately 8 s. Each truncated speech segment is generated by randomly splitting one testing speech recording into speech segments with durations of 2 s, 3 s or 5 s. These truncated speech segments in the Vox1-O, Vox1-E and Vox1-H, are used to assess the robustness of different methods to the duration of truncated speech segments. Table 7 presents experimental results obtained by different methods on the truncated testing speech segments, in which "whole" stands for the complete testing utterance without truncation.

**Table 7**

EER (in %) and minDCF obtained by different methods on truncated testing speech segments with different durations.

| Methods | Segment duration | Vox1-O (EER/minDCF) | Vox1-E (EER/minDCF) | Vox1-H (EER/minDCF) |
|---|---|---|---|---|
| TDNN based (PLDA) | 2 s | 6.208/0.631 | 6.634/0.671 | 9.024/0.895 |
| | 3 s | 3.692/0.398 | 4.118/0.427 | 6.487/0.663 |
| | 5 s | 3.156/0.338 | 3.398/0.365 | 5.310/0.561 |
| | Whole | 2.883/0.312 | 3.032/0.326 | 4.891/0.528 |
| ResNet18 based (CSM) | 2 s | 4.041/0.432 | 4.428/0.448 | 6.644/0.689 |
| | 3 s | 2.968/0.319 | 3.015/0.365 | 5.282/0.537 |
| | 5 s | 2.245/0.244 | 2.433/0.267 | 4.123/0.462 |
| | Whole | 1.936/0.225 | 2.120/0.247 | 3.794/0.413 |
| ResNet34 based (CSM) | 2 s | 3.728/0.416 | 3.941/0.423 | 6.165/0.637 |
| | 3 s | 2.496/0.274 | 2.502/0.288 | 4.827/0.482 |
| | 5 s | 1.847/0.221 | 1.949/0.237 | 3.843/0.408 |
| | Whole | 1.674/0.193 | 1.752/0.195 | 3.440/0.389 |
| Res2Net based (CSM) | 2 s | 3.697/0.407 | 3.894/0.414 | 6.111/0.615 |
| | 3 s | 2.445/0.268 | 2.468/0.281 | 4.774/0.451 |
| | 5 s | 1.803/0.213 | 1.906/0.235 | 3.801/0.355 |
| | Whole | 1.632/0.187 | 1.714/0.194 | 2.894/0.301 |
| LSTMN based (CSM) | 2 s | 4.281/0.471 | 4.413/0.498 | 6.673/0.683 |
| | 3 s | 2.997/0.337 | 3.024/0.362 | 5.248/0.525 |
| | 5 s | 2.433/0.282 | 2.412/0.303 | 4.413/0.449 |
| | Whole | 2.245/0.267 | 2.288/0.275 | 3.966/0.441 |
| GAFAM based (CSM) | 2 s | 3.805/0.427 | 4.019/0.438 | 6.306/0.651 |
| | 3 s | 2.573/0.283 | 2.580/0.303 | 4.969/0.495 |
| | 5 s | 1.922/0.231 | 2.027/0.252 | 3.985/0.421 |
| | Whole | 1.751/0.212 | 1.829/0.209 | 3.581/0.402 |
| AMCRN based (CSM) | 2 s | 3.254/0.349 | 3.240/0.337 | 4.792/0.493 |
| | 3 s | 2.159/0.240 | 1.818/0.219 | 3.607/0.372 |
| | 5 s | 1.600/0.185 | 1.665/0.193 | 2.816/0.337 |
| | Whole | 1.464/0.177 | 1.551/0.182 | 2.640/0.284 |

Based on the results achieved by different methods in Table 7, the following four observations can be obtained.

First, the scores of both EER and minDCF of all methods on all testing data subsets steadily increase with the decrease of the durations of testing speech segments. In addition, the increase of both EER and minDCF scores obtained by our method is less than that of the baseline methods. For instance, when the durations of speech segments in the Vox1-H decrease from 5 s to 2 s, the absolute increasement of the EER score obtained by our method is 1.976% (4.792% - 2.816%). However, the counterparts obtained by the baseline methods are not less than 2.260% (6.673% - 4.413%).

Second, the shorter the testing speech segment, the larger the scores of both EER score and minDCF score. For example, the EER scores of our method increase from 2.640% to 4.792% when the durations of testing speech segments in the Vox1-H decrease from "Whole" (approximately 8 s) to 2 s. The similar results can be obtained for the baseline methods.

Third, our method obtains the lowest scores of both EER and minDCF, whereas the TDNN based (PLDA) method achieves the highest scores of both EER and minDCF on all testing data subsets. That is, our method outperforms the baseline methods in terms of both EER and minDCF when they are evaluated on truncated speech segments with various durations.

Fourth, the shorter the duration of testing speech segment is, the more significant the performance margins between our method and the baseline methods are. For example, the absolute margin of EER score between our method and the TDNN based (PLDA) method is 2.251% (4.891% - 2.640%) when the speech segments with duration of "Whole" in the Vox1-H are adopted. However, the counterpart between these two methods becomes 4.232% (9.024% - 4.792%) when the speech segments with duration of 2 s are used as testing data.

In summary, our method still outperforms the baseline methods when they are evaluated on the truncated testing speech segments with different durations. In addition, compared with the baseline methods, our method is more robust to the duration of truncated speech segments. The possible reason is that the speaker embedding learned by the proposed AMCRN can effectively represent global sequential information, local spatial information with many resolutions, and the characteristic differences among different speakers from critical speech frames. In contrast, the speaker embeddings learned by the baseline deep models don't have such abilities. As a result, the proposed method can generalize well across truncated speech segments with different durations instead of overfitting the speech segments with one single duration.

*3.6. Computational complexity and memory requirement of different methods*

In this sub-section, we compare our method with the baseline methods in terms of both computational complexity and memory requirement. The computational complexity is measured by the metric of Multiply-Accumulate operations (MACs), while the memory requirement is measured by the metric of Model Size (MS). MACs stands for the number of multiplication and addition operations of a deep model. MS denotes total number of all parameters of a deep model. The lower the score of MACs is, the lower computational complexity of the method is. Similarly, the lower the score of MS is, the lower memory requirement of the method is. The computational complexity and memory requirement of different methods are mainly determined by the network (e.g., TDNN) which is used for learning speaker embedding. The reason is that all methods use the same back-end classifier for scoring, except the TDNN based (PLDA) method. In addition, the computational complexity and memory requirement of different networks adopted in various methods mainly depend on the structure and the number of blocks (or layers) of these networks.

The computational complexities and memory requirements of different methods are presented in Table 8. In terms of memory requirement, the MS of our model (AMCRN) is 11.4 M (Million) which is smaller than that of ResNet18 and ResNet34, but is larger than that of TDNN, Res2Net, LSTMN, and GAFAM. In terms of computational complexity, the MACs of our method are 0.56 G (Giga) on speech segments with 2 s, 0.84 G on speech segments with 3 s, and 1.39 G on speech segments with 5 s.

Furthermore, the MACs of our method are lower than the counterparts of all baseline methods when they are evaluated on speech segments with different durations. In summary, our method has advantage over the baseline methods of both ResNet18 based (CSM) and ResNet34 based (CSM), but lacks advantage over other four baseline methods in terms of memory requirement. In addition, our method has advantage over all baseline methods in terms of computational complexity.

**Table 8**

Computational complexities and memory requirements of different methods.

| Methods | MS | MACs | | |
|---|---|---|---|---|
| | | 2 s | 3 s | 5 s |
| TDNN based (PLDA) | 7.68 M | 0.57 G | 0.85 G | 1.41 G |
| ResNet18 based (CSM) | 13.8 M | 4.82 G | 7.15 G | 11.88 G |
| ResNet34 based (CSM) | 23.9 M | 10.07 G | 14.95 G | 24.83 G |
| Res2Net based (CSM) | 10.6 M | 1.34 G | 2.12 G | 3.21 G |
| LSTMN based (CSM) | 8.6 M | 0.58 G | 0.92 G | 1.48 G |
| GAFAM based (CSM) | 9.8 M | 3.42 G | 5.08 G | 8.44 G |
| AMCRN based (CSM) | 11.4 M | 0.56 G | 0.84 G | 1.39 G |

## 4. Conclusions

In this study, we propose a speaker verification method by a deep model of the proposed AMCRN. The SEM contained in the AMCRN is designed to learn speaker embedding. Based on the description of the proposed method and the experiments, we can draw the following three conclusions.

First, the proposed AMCRN is an effective deep model for learning speaker embedding which can effectively represent the differences of time-frequency characteristics among different speakers. The speaker embedding learned by the proposed AMCRN exceeds the counterparts learned by other deep models for speaker verification. In addition, the proposed speaker embedding generalizes well across the back-end classifiers of CSM and PLDA.

Second, when evaluated on three testing datasets, the proposed method outperforms the baseline methods based on the TDNN, ResNet18, ResNet34, Res2Net, LSTMN and GAFAM, in terms of EER and minDCF. What's more, the proposed method generalizes well across truncated speech segments with various durations.

Third, the computational complexity and memory requirement of the proposed method are lower than that of the methods based on the ResNet18 and ResNet34. Compared to the methods based on the TDNN, Res2Net, LSTMN and GAFAM, our method has advantage in terms of computational complexity, but lacks advantage in terms of memory requirement.

The future work includes two parts. First, we will investigate the structures of main blocks of the AMCRN to further decrease the scores of EER and minDCF, such as convolutional-capsule block, and self-attention block. Second, to meet the requirements of lightweight applications, we will reduce the computational complexity and memory requirement of the proposed method by further optimization, such as using depth-wise separable convolution and network pruning techniques. As a result, we can implement our method on portable terminals with limited computational resources.

## CRediT authorship contribution statement

**Yanxiong Li**: Conceptualization, Methodology, Formal analysis, Investigation, Writing - original draft, Writing - review & editing, Validation, Supervision, Project administration. **Zhongjie Jiang**: Conceptualization, Methodology, Writing - original draft, Writing - review & editing, Validation.

**Wenchang Cao**: Conceptualization, Methodology, Software, Validation, Writing - original draft.
**Qisheng Huang**: Software, Validation, Writing - review & editing.

## Declaration of Competing Interest

The authors declare that they have no known competing financial interests or personal relationships that could have appeared to influence the work reported in this paper.

## Acknowledgements

This work was partly supported by national natural science foundation of China (62111530145, 61771200), Guangdong basic and applied basic research foundation, China (2021A1515011454), and international scientific research collaboration project of Guangdong Province, China (2021A0505030003).